\documentclass[reprint, aps, pra, letterpaper, superscriptaddress]{revtex4-2}%

\usepackage[T1]{fontenc} 

\usepackage{amsmath,color}
\usepackage{amssymb}
\usepackage{amsfonts}
\usepackage{amsthm}
\usepackage{mathtools}

\usepackage[title]{appendix}

\usepackage{algorithm}
\usepackage{algpseudocode}
\usepackage{dsfont}

\usepackage{bbold}
\usepackage{chemformula}
\usepackage{siunitx}

\DeclarePairedDelimiterX\braket[2]{\langle}{\rangle}{#1 \delimsize\vert #2}
\DeclarePairedDelimiterX\braket3[3]{\langle}{\rangle}{#1 \delimsize\vert #2 \delimsize\vert #3}

\usepackage{accents}
\newcommand{\dbtilde}[1]{\accentset{\approx}{#1}}

\newcommand{\ve}{\mathbf{e}}

\newcommand{\vmu}{\boldsymbol{\mu}}

\newcommand{\kB}{k_{\text{B}}}

\newcommand{\avg}[1]{\left\langle #1\right\rangle}

\begin{document}

	\title{Collective vibrational strong coupling effects on molecular vibrational relaxation and energy transfer: Numerical insights via cavity molecular dynamics simulations}
	
	\author{Tao E. Li}%
	\email{taoli@sas.upenn.edu}
	\affiliation{Department of Chemistry, University of Pennsylvania, Philadelphia, Pennsylvania 19104, USA}

	\author{Abraham Nitzan} 
	\email{anitzan@sas.upenn.edu}
	\affiliation{Department of Chemistry, University of Pennsylvania, Philadelphia, Pennsylvania 19104, USA}
	\affiliation{School of Chemistry, Tel Aviv University, Tel Aviv 69978, Israel}

	\author{Joseph E. Subotnik}
	\email{subotnik@sas.upenn.edu}
	\affiliation{Department of Chemistry, University of Pennsylvania, Philadelphia, Pennsylvania 19104, USA}
	
	\begin{abstract}
		For a small fraction of hot \ch{CO2} molecules immersed in a liquid-phase \ch{CO2} thermal bath, classical cavity molecular dynamics simulations show that forming collective vibrational strong coupling (VSC) between the \ch{C=O} asymmetric stretch of  \ch{CO2} molecules and a cavity mode accelerates hot-molecule relaxation.
		The physical mechanism underlying this acceleration is the fact that polaritons, especially the lower polariton, can be transiently excited during the nonequilibrium process, which facilitates intermolecular vibrational energy transfer. The VSC effects on these rates (i) resonantly depend on the cavity mode detuning, (ii) cooperatively depend on molecular concentration or Rabi splitting, and (iii) collectively scale  with the number of hot molecules, which is similar to Dicke's superradiance. For  larger cavity volumes, due to a balance between this superradiant-like behavior and a smaller light-matter coupling, the  total VSC effect on relaxation rates can scale slower than $1/N$, and the \textit{average} VSC effect \textit{per molecule} can remain meaningful for up to $N \sim10^4$ molecules forming VSC. Moreover, we find that the transiently excited lower polariton prefers to relax by transferring its energy to the \textit{tail} of the molecular energy distribution rather than equally distributing it to all thermal molecules. 
		Finally, we highlight the similarities of parameter dependence between the current finding with VSC catalysis observed in Fabry--P\'erot microcavities.
	\end{abstract}
	
	\maketitle
	
	\section{Introduction}\label{sec:intro}
	
	Collective vibrational strong coupling (VSC) can occur if a macroscopic number of liquid-phase molecules are confined to a Fabry--P\'erot  microcavity and a molecular vibrational mode is near resonant with a cavity mode \cite{George2015,George2016}. Under collective VSC, experimental reports indicate not only a peak splitting, i.e., a Rabi splitting within molecular infrared (IR) spectroscopy,  but also the modification of chemical reaction rates \cite{Thomas2016,Thomas2019,Thomas2019_science,Lather2019} and crystallization processes \cite{Hirai2020} under thermal conditions.   As pioneered first by Ebbesen \cite{Thomas2016} and co-workers, these observations suggest that collective VSC might meaningfully modify individual molecular properties \textit{without} external pumping ---
	although these intriguing experimental findings cannot yet be well explained by current theory  \cite{Galego2019,Campos-Gonzalez-Angulo2019,Li2020Origin,Campos-Gonzalez-Angulo2020,Sidler2021,LiHuo2021}.
	
	A simple example illustrating how conventional theory fails to explain the Ebbesen experiments is to consider the case of $N$ molecules forming VSC with a Rabi splitting $\Omega_N = 2g_0\sqrt{N} \sim 100$ cm$^{-1}$, where $g_0$ denotes the light-matter coupling for individual molecules. Because $g_0$ ($=\Omega_N /2\sqrt{N}$) is negligible when $N$ becomes macroscopic, one would guess that individual molecular properties (such as chemical reaction rates) cannot be meaningfully modified by a Fabry--P\'erot  microcavity, a theoretical prediction at odds with several experiments. Recent efforts \cite{Li2020Origin,Li2020Water} also suggest that, within a classical description of cavity photons and molecules, static properties of individual molecules during thermal equilibrium are entirely unchanged under usual VSC setups, indicating a nonequilibrium (or perhaps quantum) origin of the Ebbesen experiments.
	
	In order to narrow the gap between theory and experiment, here we numerically investigate VSC effects on two nonequilibrium processes ---  molecular vibrational energy relaxation and intermolecular vibrational energy transfer.   These vibrational processes has been extensively studied both experimentally and theoretically outside a cavity, and the rates of which have been known to play an important role in many physical and chemical processes, including chemical reactions \cite{Gruebele2004}. Inside a cavity, a recent experiment \cite{Xiang2020Science} has studied the effect of VSC on intermolecular vibrational energy transfer rates by quantifying the response of hybrid light-matter states (polaritons) after pumping the upper polariton (UP) for  a liquid mixture of \ch{W($^{12}$CO)6} and \ch{W($^{13}$CO)6}.

	For the sake of simplicity, here our numerical study focuses on a pure  liquid \ch{CO2} system when the \ch{C=O} asymmetric stretch forms VSC with a single optical cavity mode (where two polarization directions are included). In such a system, instead of exciting polaritons, we will consider the case when a small fraction of uncorrelated hot \ch{CO2} molecules dissipates and transfers vibrational energy to the remaining thermal \ch{CO2} molecules at room temperature. 
	Unlike many experiments and theoretical studies concentrating on the polaritonic response, we will mainly focus on how individual molecules (which are mostly composed of vibrational dark modes) relax and transfer energy under VSC. In detail, we will extensively study how vibrational energy relaxation and  transfer depend on cavity mode detuning, molecular concentration (or Rabi splitting), and the number of hot molecules. Because there is no external polariton pumping, our investigation of   how a cavity affects vibrational relaxation and energy transfer will hopefully yield insight into  the VSC modifications of individual molecular properties (such as chemical reaction rates) that are observed in experiments. In particular, by quantifying the asymptotic scaling of VSC effects  with molecular system size or effective cavity volumes, our study will also partly address if  VSC effects can persist and affect  the properties of  individual molecules in the limit that a very large number of molecules are present in a cavity.

	The theoretical approach we will take is classical cavity molecular dynamics (CavMD) simulations \cite{Li2020Water,Li2020Nonlinear}, a newly developed numerical tool implemented by the authors to classically propagate the coupled dynamics between realistic molecules (assumed to stay in their electronic ground-state) and  cavity photons in the dipole gauge. Since the self-dipole term is included in the light-matter Hamiltonian of CavMD simulations, this numerical approach preserves gauge invariance and maintains numerical stability \cite{Schafer2020}. Compared with VSC experiments, this approach has reliably captured many VSC-induced phenomena, including an asymmetric Rabi splitting \cite{Li2020Water,Vergauwe2019}, polariton relaxation to vibrational dark modes on a time scale of ps and sub-ps \cite{Li2020Nonlinear,Xiang2018}, and a delay of population gain in the  singly excited manifold of vibrational dark modes after pumping the lower polariton (LP), a process which stems from polariton enhanced molecular nonlinear absorption \cite{Li2020Nonlinear,Xiang2019State}.  Hence, CavMD simulations appears to be a promising tool to study VSC-related phenomena.

	\begin{figure}
		\centering
		\includegraphics[width=1.0\linewidth]{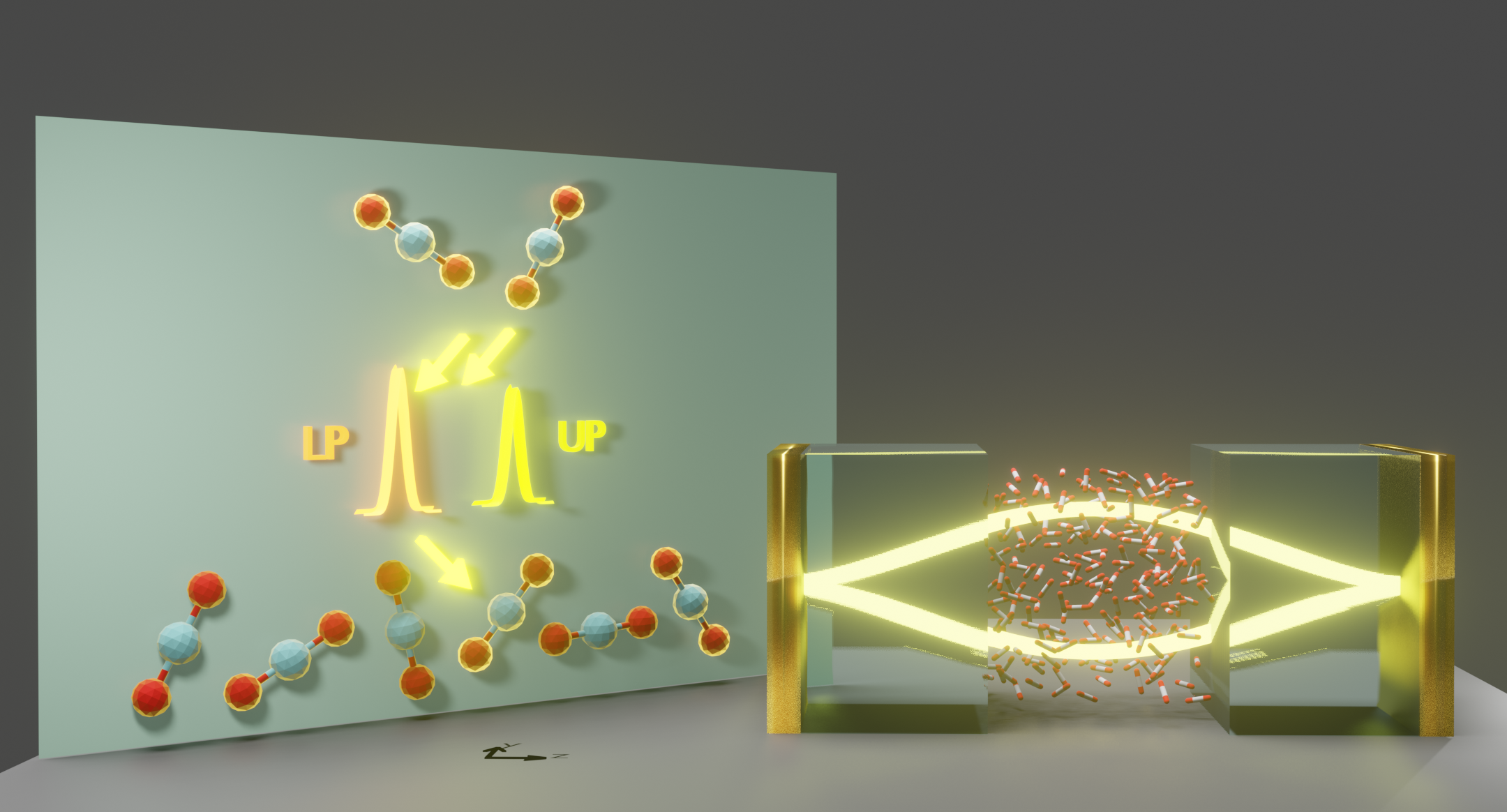}
		\caption{Sketch of the simulation setup where a large number of  liquid-phase carbon dioxide molecules forms VSC with a single cavity mode. The left cartoon shows that the vibrational energy relaxation and transfer from hot (upper) to thermal (bottom) molecules inside a cavity can be accelerated relative to that outside a cavity due to polariton-accelerated intermolecular vibrational energy transfer.}
		\label{fig:toc}
	\end{figure}

 	A brief introduction of CavMD is given in Appendix \ref{sec:method} and \ref{sec:simulation_details};  see Ref. \cite{Li2020Nonlinear} for more details regarding CavMD simulations  of a liquid \ch{CO2} system and how the \ch{CO2} force field is defined.  In short,  as shown in Fig. \ref{fig:toc}, CavMD simulates a system with $N_{\text{sub}}$ \ch{CO2} molecules in a periodic cell coupled to a single cavity mode (with two polarization directions $x$ and $y$). The effective coupling strength between each molecule and the cavity mode is $\widetilde{\varepsilon}$. Note that, during nonequilibrium CavMD simulations, we have disregarded cavity loss. This simplification is valid because in Fabry--P\'erot microcavities, the dominant  channel for polaritons to relax is through vibrational dark modes (with a lifetime of ps or sub-ps with our parameter setting \cite{Li2020Nonlinear}) and cavity loss takes a longer lifetime ($\sim 5$ ps). Below, we will report how VSC affects vibrational energy relaxation and transfer using CavMD simulations.

	\section{Results}\label{sec:results}

	\subsection{VSC effects on vibrational energy relaxation and transfer}
	
	\begin{figure*}
		\centering
		\includegraphics[width=1.0\linewidth]{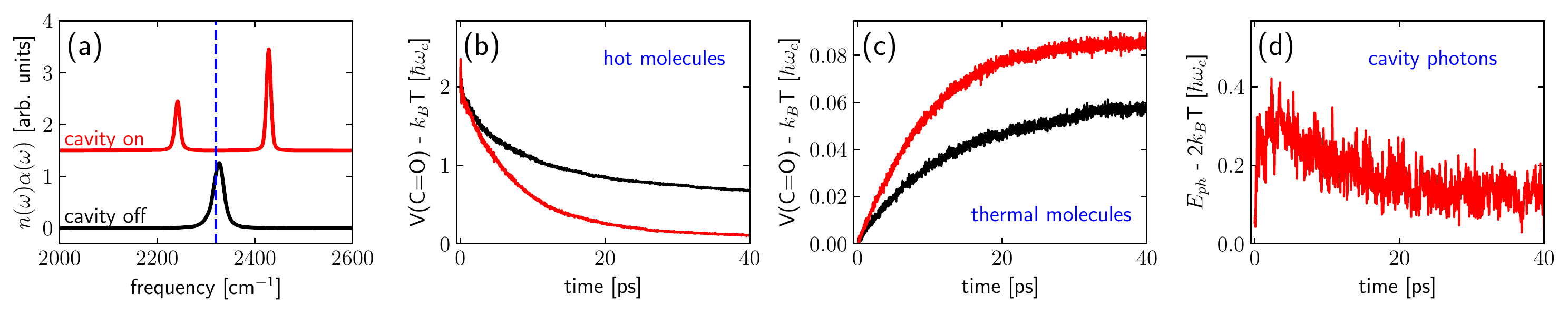}
		\caption{Rabi splitting and VSC effects on vibrational energy relaxation and transfer when $N_{\text{sub}} = 216$ and $N_{\text{hot}} = 10$. (a) Simulated IR spectrum for liquid \ch{CO2} outside (black) or inside (red) the cavity. For parameters, the cavity mode frequency is set to $\omega_c=2320$ cm$^{-1}$ (denoted as the  vertical blue line) and the effective coupling strength $\widetilde{\varepsilon} = 2\times 10^{-4}$ a.u. (inside the cavity) or zero (outside the cavity). (b,c) The corresponding average \ch{C=O} bond  potential energy (per molecule) dynamics for the (b) hot or (c) thermal molecules   outside (black) or inside (red) the cavity. (d) The corresponding photonic (kinetic + potential) energy dynamics inside the cavity, where two polarization directions of the cavity mode are taken into account.  In the $y$-axis of Figs. b-d, a thermal energy (i.e., $k_BT$ for Figs. b,c and $2k_B T$ for Fig. d) has been subtracted and all energies are in units of $\hbar\omega_c$. See Appendix \ref{sec:simulation_details} for other simulation details. Note that polaritons play an important role during the process of vibrational energy relaxation and transfer as evidenced from the high transient photonic energy as compared with the vibrational energy transferred to the thermal molecules.}
		\label{fig:cavity_effect}
	\end{figure*}

	 Fig. \ref{fig:cavity_effect}a plots the IR spectrum   outside the cavity (black line; $\widetilde{\varepsilon} = 0$) or inside the cavity (red line; $\widetilde{\varepsilon} = 2\times 10^{-4}$ a.u.)   when the cavity mode (at $\omega_c = 2320$ cm$^{-1}$; the dashed vertical blue line) forms VSC with the \ch{C=O} asymmetric mode (peaked at $\omega_0 = 2327$ cm$^{-1}$) of liquid \ch{CO2}. 
	Inside the cavity,  a pair of lower (LP; peaked at 2241 cm$^{-1}$) and upper (UP; peaked at 2428 cm$^{-1}$) polaritons form and these polaritons are separated by a Rabi splitting of 187 cm$^{-1}$. The IR spectrum is calculated by evaluating the Fourier transform of the dipole autocorrelation function from equilibrium trajectories; see Appendix \ref{Appendix:spectrum} for details.

	We now consider a nonequilibrium process where $N_{\text{hot}}=10$ uncorrelated hot molecules are immersed in a thermal \ch{CO2} bath at room temperature (where in total there are $N_{\text{sub}} = 216$ molecules in the simulation cell); see Appendix \ref{sec:simulation_details} for details. Fig. \ref{fig:cavity_effect}b plots the average time-resolved \ch{C=O} bond potential energy per hot molecule outside (black line) or inside (red line) the cavity, where a thermal energy $k_B T = 300$ K has been subtracted from the \ch{C=O} bond potential energy; note that here we use $k_BT$ instead of $k_BT/2$ since each \ch{CO2} molecule contains two \ch{C=O} bonds. As shown in Fig. \ref{fig:cavity_effect}b, the initial potential energy in the  two \ch{C=O} bonds per hot molecule is roughly $2\hbar\omega_c \ (\approx 6\times 10^3 \text{\ K})$, i.e., the initial temperature of the hot molecules is $\sim 3\times 10^3$ K.
	At later times, the vibrational energy relaxation of the hot molecules inside the cavity is accelerated compared with that outside the cavity.
	Meanwhile, as shown in Fig. \ref{fig:cavity_effect}c, the average \ch{C=O} bond potential energy per thermal molecule   inside (red line) the cavity increases faster than that outside (black line) the cavity. Here,  "thermal molecules" refer to molecules that were prepared at thermal equilibrium. During this nonequilibrium process, the total system energy is  conserved:  the simulation is performed under a NVE (constant number, volume, and energy) ensemble; see simulation details in Appendix \ref{sec:method}. 
	
	During the energy relaxation and transfer process, inside the cavity, Fig. \ref{fig:cavity_effect}d plots the total (kinetic + potential) energy of the cavity photons ($\omega_c = 2320$ cm$^{-1}$ and with two polarization directions) subtracted by the thermal background $2k_B T$.  Because cavity photons contribute half of the polaritons, Fig. \ref{fig:cavity_effect}d  indicates that polaritons can be transiently excited during this nonequilibrium process. 
	Note that, at long times, the cavity photon energy does not decay back to zero because the relaxation of the hot molecules will  increase the system temperature to above 300 K. From Figs. \ref{fig:cavity_effect}b-c, we can conclude that the cavity acceleration of vibrational energy relaxation stems from  cavity-accelerated intermolecular vibrational energy transfer from the hot  to the thermal molecules. Furthermore, compared with thermal molecules (see Fig. \ref{fig:cavity_effect}c red line), cavity photons can be excited more meaningfully at short times. This fact emphasizes the importance of forming polaritons and the interaction between polaritons and individual molecules (which are predominately composed of vibrational dark modes) in modifying these rates.

	\subsection{Detuning dependence}
	
	\begin{figure}
		\centering
		\includegraphics[width=1.0\linewidth]{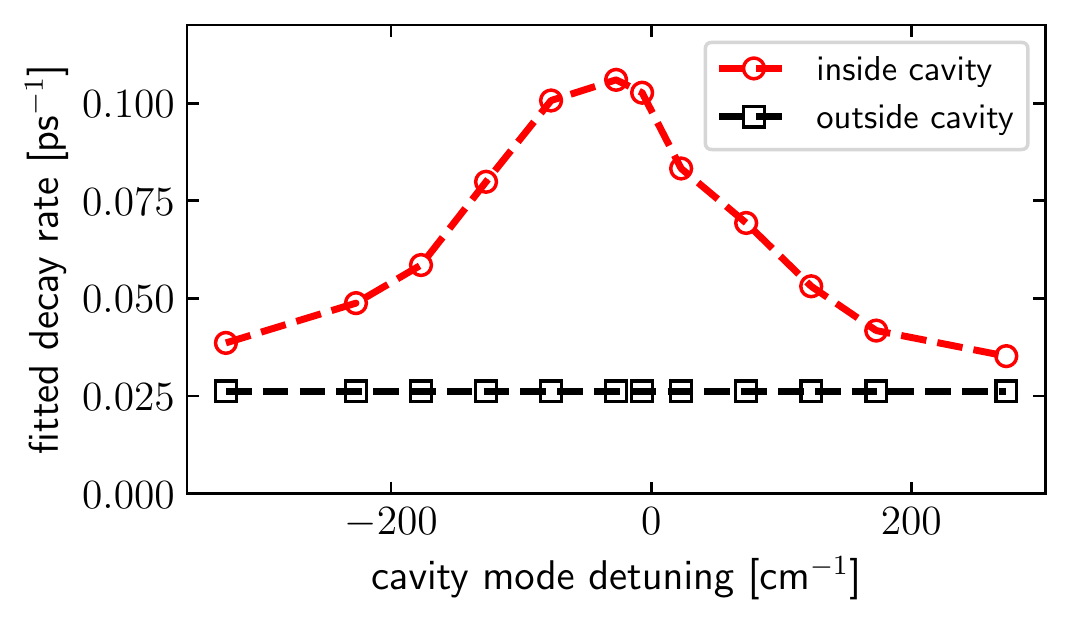}
		\caption{Fitted vibrational energy relaxation rates as a function of the cavity mode detuning. All parameters are the same as in Fig. \ref{fig:cavity_effect}b except that we now change the cavity mode frequency ($\omega_c$). Rates are obtained by fitting the signals in Fig. \ref{fig:cavity_effect}b to a simple exponential function: $y = A\exp(-kt)$. Note that the VSC effect on vibrational energy relaxation resonantly depends on the cavity mode frequency.}
		\label{fig:detuning_dependence}
	\end{figure}
	
	Consider now that case where the cavity photon frequency is changed but all other parameters are the same as in Fig. \ref{fig:cavity_effect}. Fig. \ref{fig:detuning_dependence} plots the fitted vibrational energy relaxation rates (using an exponential function $y = A\exp(-kt)$ to fit Fig. \ref{fig:cavity_effect}b) of the hot molecules against the cavity mode detuning $\delta = \omega_c - \omega_0$. Compared with the fitted decay rates outside the cavity (black squares), the rates inside the cavity (red circles) show a resonant dependence on the detuning $\delta$: when $\delta \approx 0$, the maximum rate inside the cavity is roughly four times the rate outside the cavity. Because the cavity mode is decoupled from \ch{C=O} asymmetric stretch under a large detuning, this resonance behavior again points to the importance of forming polaritons as far as modifying relaxation rates. We have also found (not shown here) that  the VSC effect on relaxation rate depends only weakly on the temperature of the hot molecules.

	\subsection{Rabi splitting dependence}
	\begin{figure*}
		\centering
		\includegraphics[width=0.8\linewidth]{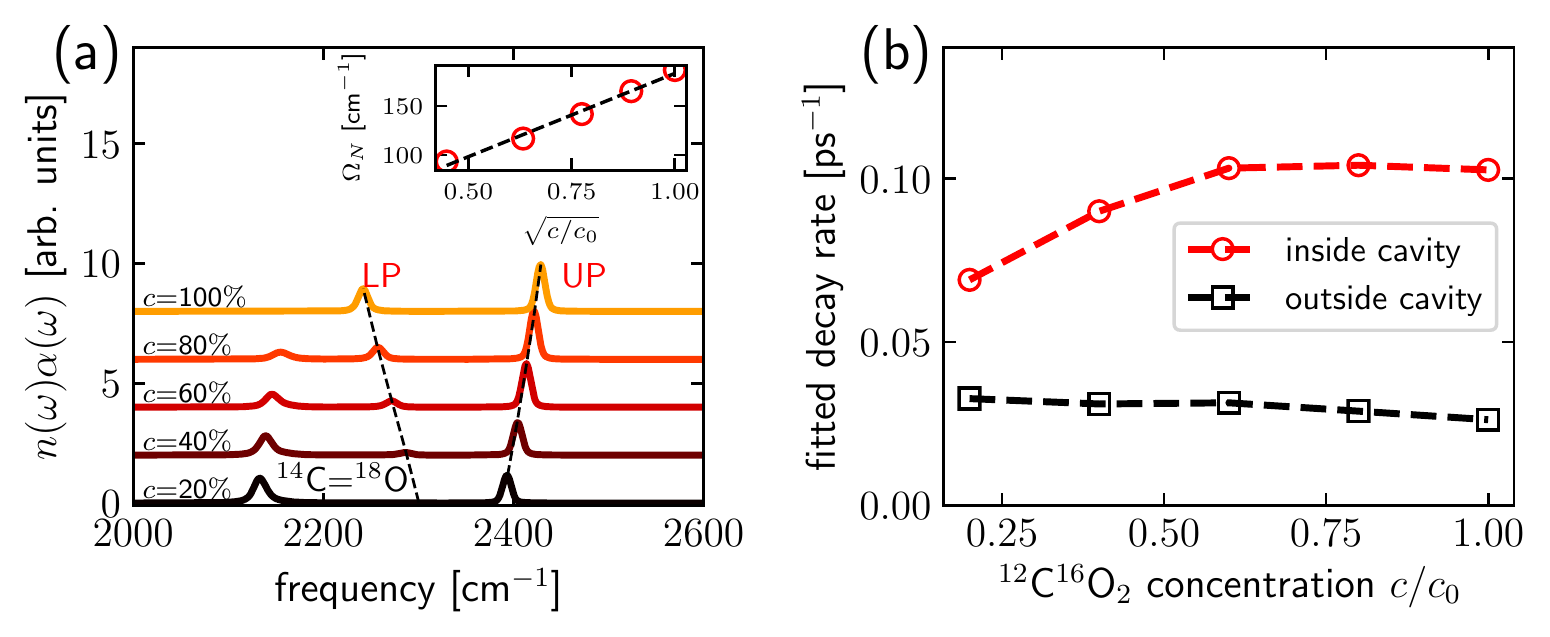}
		\caption{VSC effects on vibrational energy relaxation as a function of Rabi splitting. (a) Simulated IR spectrum for a liquid mixture of \ch{$^{12}$C$^{16}$O2} and \ch{$^{14}$C$^{18}$O2} inside the cavity. The cavity mode ($\omega_c=2320$ cm$^{-1}$) forms polaritons with the \ch{$^{12}$C=$^{16}$O} asymmetric stretch mode ($\omega_0=2327$ cm$^{-1}$), while the \ch{$^{14}$C=$^{18}$O} asymmetric stretch mode (the leftest peak) is largely decoupled from the cavity. Note that when the \ch{$^{12}$C$^{16}$O2} concentration increases from $c=20$\% to 100\% (bottom to to top), the Rabi splitting ($\Omega_N$) is also increased proportionally; see the inset. (b) The corresponding fitted vibrational relaxation rates for the hot \ch{$^{12}$C$^{16}$O2} molecules plotted against the \ch{$^{12}$C$^{16}$O2} concentration ($c/c_0$), where $c_0 = 100\%$ denotes the concentration for a pure \ch{$^{12}$C$^{16}$O2} system. The rates inside the cavity (red circles) show a sensitive dependence on the \ch{$^{12}$C$^{16}$O2} concentration (or Rabi splitting), while the rates outside the cavity (black squares) show a weak dependence on the concentration.}
		\label{fig:concentration_dependence}
	\end{figure*}
	
	We next investigate how vibrational energy relaxation rates depend on the Rabi splitting by introducing an isotopic liquid mixture of carbon dioxide and  changing the relative molecular concentration of each isotopic form.
	With all other parameters the same as Fig. \ref{fig:cavity_effect} (where a pure \ch{CO2}, or \ch{$^{12}$C$^{16}$O2} system is studied), Rabi splitting is tuned by replacing some \ch{$^{12}$C$^{16}$O2} molecules by \ch{$^{14}$C$^{18}$O2}.
	 Fig. \ref{fig:concentration_dependence}a plots the equilibrium IR spectrum inside the cavity under an increased concentration of \ch{$^{12}$C$^{16}$O2} ($c = 20\%$ to $100\%$ from bottom to top). Because \ch{$^{14}$C$^{18}$O2} is relatively heavy, the \ch{$^{14}$C=$^{18}$O} asymmetric stretch (the leftest peak in Fig. \ref{fig:concentration_dependence}a) is well separated from the \ch{$^{12}$C=$^{16}$O} asymmetric stretch (peaked at $\omega_0 = 2327$ cm$^{-1}$), and \ch{$^{14}$C$^{18}$O2}  molecules effectively do not participate in the formation of polaritons (LP and UP in Fig. \ref{fig:concentration_dependence}a) between the cavity mode  (peaked at $2320$ cm$^{-1}$) and the \ch{$^{12}$C=$^{16}$O} asymmetric stretch. The inset of Fig. \ref{fig:concentration_dependence}a plots the Rabi splitting $\Omega_N$ between the UP and LP as a function of $\sqrt{c/c_0}$, where $c_0 = 100\%$ denotes the concentration of pure \ch{$^{12}$C$^{16}$O2}. As in many experiments, a linear scaling between $\Omega_N$ and $\sqrt{c/c_0}$ is observed.
	
	Under different concentrations of \ch{$^{12}$C$^{16}$O2}, Fig. \ref{fig:concentration_dependence}b plots the fitted vibrational energy relaxation rates when  10 hot \ch{$^{12}$C$^{16}$O2} molecules  ($N_{\text{hot}}=10$)  are immersed in the liquid mixture.  The outside-cavity results (black squares) show  a weak dependence on the molecular concentration. 
	 By contrast, inside the cavity (red circles), we observe an obvious acceleration of the relaxation rates when the \ch{$^{12}$C$^{16}$O2} concentration is increased from $c=20\%$ to $60\%$ and then a plateau region above  $c=60\%$. This acceleration of the relaxation rates (with a monotonic dependence on molecular concentration) shows that, inside a cavity, the relaxation of a few molecules indeed depends strongly on the total molecular number (or concentration).
	 
	Interestingly, experiments outside a cavity  \cite{Shaw2009} have shown that vibrational relaxation rates in hydrogen-bonded liquids (\ch{X-H}/\ch{X-D} mixture)  demonstrate similar sensitive dependence on isotope concentration as we have found  inside a cavity in Fig. \ref{fig:concentration_dependence}b. In Ref. \cite{Shaw2009}, the authors argued that such isotopic dependence can be explained by noting that, for a system with hydrogen bonding, intermolecular  vibrational energy transfer can be facilitated by forming a delocalized intermediate  state between two neighboring molecules. In an analogous pattern, Fig. \ref{fig:concentration_dependence}b implies that polaritons can similarly serve as a "delocalized intermediate state" and facilitate intermolecular vibrational energy transfer even in weakly interacting liquids.

	\subsection{Superradiant-like collective relaxation}\label{sec:collective}
	
	After demonstrating that VSC leads to cooperative effects on vibrational energy relaxation rates against the molecular concentration or Rabi splitting,  we next study how vibrational relaxation rates depend on the number of {\em hot} molecules ($N_{\text{hot}}$). Going beyond Fig. \ref{fig:cavity_effect} (where $N_{\text{sub}} = 216$ molecules are confined in a periodic simulation cell), here we simulate $N_{\text{sub}} = 2160$ molecules  while keeping all other macroscopic variables --- such as molecular density (1.101 g/cm$^3$) and the Rabi splitting --- unchanged. Note that we maintain a constant Rabi splitting  by reducing the effective light-matter coupling  ($\widetilde{\varepsilon}$) for each molecule. Physically speaking, increasing the number of  molecules while adjusting the coupling so as to keeping the Rabi splitting constant corresponds to increasing the effective volume of the cavity at constant molecular density.

	\begin{figure*}
		\centering
		\includegraphics[width=0.8\linewidth]{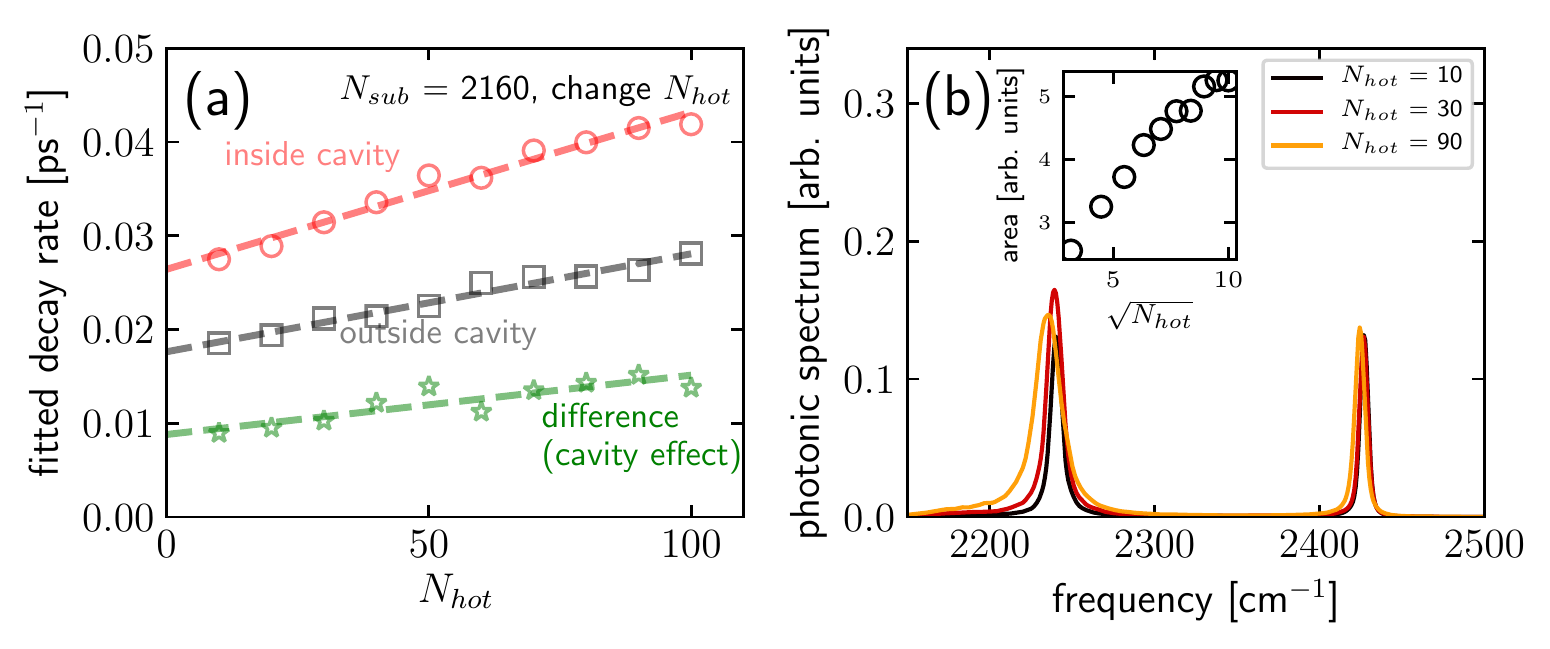}
		\caption{
		(a) Fitted vibrational energy relaxation rates for the hot molecules inside (red circles) or outside (black squares) the cavity are plotted versus $N_{\text{hot}}$, with $N_{\text{sub}}=2160$ \ch{CO2} molecules in the simulation cell. The difference between the inside- and outside-cavity results (green stars) is a pure VSC effect.Lines with different colors denote the respective linear fits of the rates. 
		(b) Corresponding photonic spectrum (which effectively represents a polaritonic spectrum) during the relaxation process when $N_{\text{hot}}$  = 10 (black),  30 (red), and 90 (orange). The inset plots the integrated area of the photonic spectrum versus $\sqrt{N_{\text{hot}}}$.  Note that the intensity of the spectrum increases monotonically as the number of hot molecules increases. See Appendices \ref{sec:simulation_details} and \ref{Appendix:spectrum} for simulation details and methods to calculate the spectrum.
	}
		\label{fig:pbc_dependence_Nhot_only}
	\end{figure*}
	
	Fig. \ref{fig:pbc_dependence_Nhot_only}a plots relaxation rates versus the number of hot molecules ($N_{\text{hot}}$) inside the large molecular system with $N_{\text{sub}} = 2160$.  
	Both the inside- (red circles) and outside-cavity (black squares) rates show a linear relationship (fitted with a linear function; see lines with respective colors) against  $N_{\text{hot}}$.
	Outside the cavity, the  linear scaling against $N_{\text{hot}}$ is understandable because increasing the number of hot molecules increases the temperature of the system, enhances intermolecular collisions and strengthens dipole-dipole interactions, all of which can lead to an acceleration of the relaxation of hot molecules. More interestingly, the inside- and outside-cavity rates show different slopes against $N_{\text{hot}}$.
	The difference between these rates is plotted with green stars and represents a pure cavity effect. This pure cavity effect scales roughly linearly against $N_{\text{hot}}$, demonstrating that polariton-accelerated vibrational energy relaxation collectively depends on $N_{\text{hot}}$.
	
	Another example demonstrating the collective behavior of vibrational relaxation is shown in Fig. \ref{fig:pbc_dependence_Nhot_only}b, where we  study the frequency distribution of the transiently excited photons (which effectively  represents the polaritons). See Appendix \ref{Appendix:spectrum} for details regarding the calculation of the polariton spectrum.  As shown in Fig. \ref{fig:pbc_dependence_Nhot_only}b, the polaritonic spectrum broadens  (especially for the LP) and red-shifts when $N_{\text{hot}}$ increases  [from $N_{\text{hot}}=10$ (black line) to $N_{\text{hot}}=90$ (orange line)]. In the inset of Fig. \ref{fig:pbc_dependence_Nhot_only}b, we show that the total, integrated intensity of the  polaritonic spectrum increases monotonically versus $\sqrt{N_{\text{hot}}}$.  The inset implies that when $N_{\text{hot}}$ increases,  the LP grows in intensity and can interact more strongly with the hot molecules in the system; the end result is  an acceleration of the hot-molecule relaxation by what one might call polariton-enhanced decay. This collective behavior is reminiscent of the Dicke's superradiance phenomenon \cite{Dicke1954}, where the spontaneous emission rates of $N$ electronic two-level systems can be collectively enhanced by a factor of $N$ when all two-level systems interact with the same electromagnetic field. Here, we observe a similar behavior  because all molecules interact with the same polaritons.

	\subsection{Asymptotic scaling of system size}
	
	Let us now address the asymptotic behavior of VSC effects for different  molecular system sizes. Here, we  change the number of molecules in the simulation cell ($N_{\text{sub}}$), while keeping  the molecular density (1.101 g/cm$^{3}$) and the Rabi splitting the same. As mentioned above, this change corresponds to investigating different effective cavity volumes.  As discussed in Appendix \ref{sec:method}, under these conditions (and especially the fixed Rabi splitting), second-order perturbative calculations suggest that the VSC effects on individual molecules  should scale  as $O(\widetilde{\varepsilon}^2)=  O(1/N_{\text{sub}})$. Below we will examine the scaling behavior for realistic CavMD simulations.
	
	\begin{figure*}
		\centering
		\includegraphics[width=0.8\linewidth]{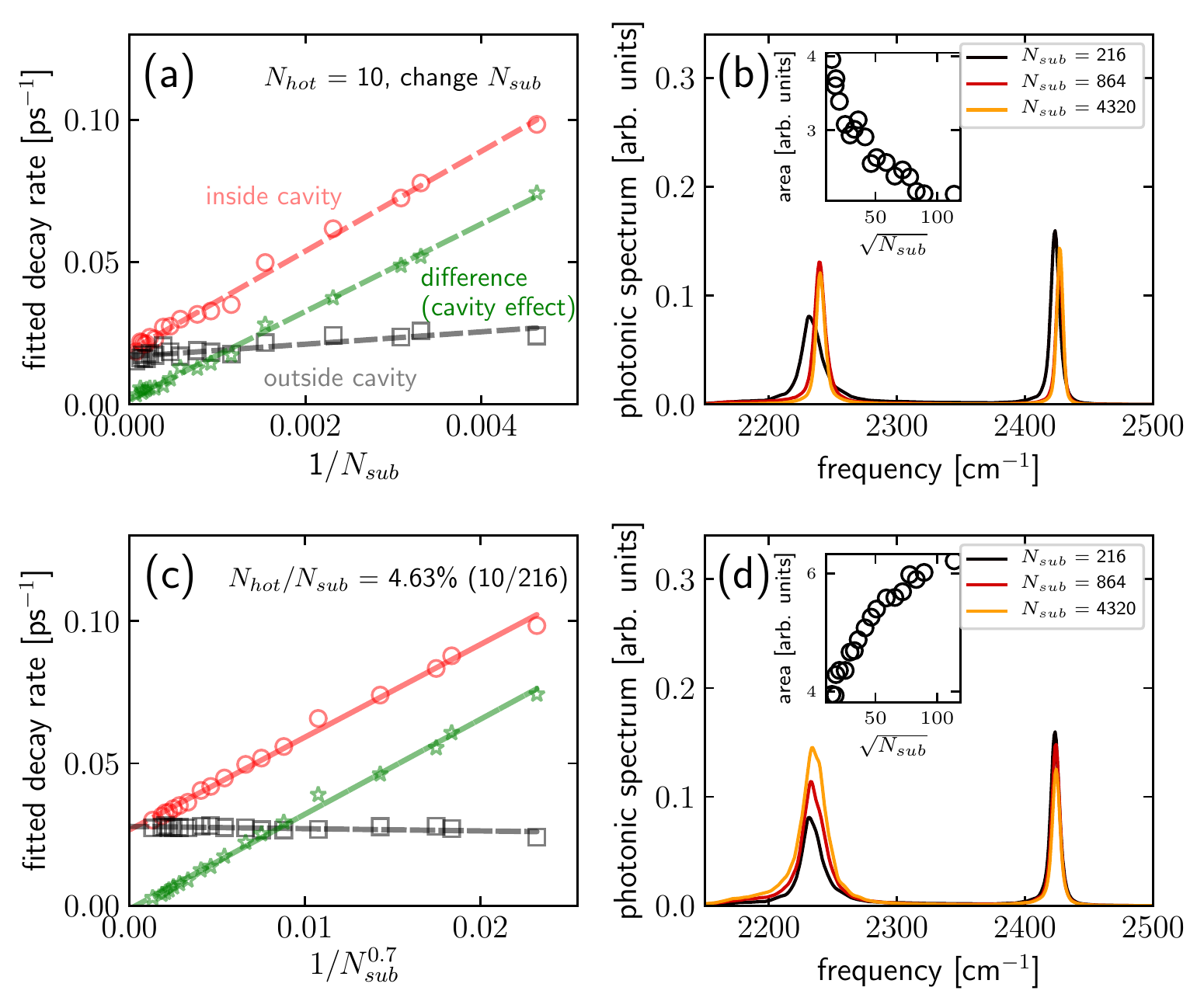}
		\caption{The dependence of  vibrational energy relaxation rates on molecular system size ($N_{\text{sub}}$).
			(a) Fitted vibrational energy relaxation rates for the hot molecules inside (red circles) or outside (black squares) the cavity against $1/N_{\text{sub}}$ when $N_{\text{hot}} = 10$ is fixed. We observe a linear scaling between the difference (green starts), which is a pure VSC effect, and $1/N_{\text{sub}}$. (b) Corresponding polaritonic spectrum for the different molecular system sizes in Fig. a: $N_{\text{sub}}$ = 216 (black), 864 (red), and 4320 (orange). The inset plots the integrated area of the photonic spectrum versus $\sqrt{N_{\text{sub}}}$.
			(c) Fitted vibrational energy relaxation rates for the hot molecules  against $1/N_{\text{sub}}^{0.7}$ when $N_{\text{hot}} / N_{\text{sub}} = 10/216 = 4.63\%$. (d) The polaritonic spectrum corresponding to  the different molecular system sizes in Fig. c.}
		\label{fig:pbc_dependence_scaling}
	\end{figure*}

	\paragraph{Standard $O(1/N_{\text{sub}})$ scaling}
	
	When the simulation system is enlarged by increasing $N_{\text{sub}}$ from 216 to 12960 and keeping the number of hot molecules ($N_{\text{hot}} = 10$) fixed,  Fig. \ref{fig:pbc_dependence_scaling}a plots the fitted average vibrational energy relaxation rates for the hot molecules inside (red circles) or outside (black squares) the cavity versus $1/N_{\text{sub}}$. 
	Outside the cavity, the rates decreases when $N_{\text{sub}}$ increases. This observation is understandable because increasing the system size while keeping $N_{\text{hot}} = 10$ effectively decreases the temperature of the system and suppresses the relaxation rate, which is consistent with the outside-cavity scaling in Fig. \ref{fig:pbc_dependence_Nhot_only}a. More importantly, the
	 difference between the inside- and outside-cavity rates --- which is a pure VSC effect --- scales linearly with  $1/N_{\text{sub}}$, which confirms the standard perturbative result. Also as shown in Fig. \ref{fig:pbc_dependence_scaling}b, when $N_{\text{sub}}$ increases, the intensity of the polaritonic spectrum decreases. This decrease arises because under a fixed Rabi splitting, the  light-matter coupling $\widetilde{\varepsilon}$ for each molecule decreases when $N_{\text{sub}}$ increases,  thus leading to a negligible polaritonic effect on the relaxation rates for the hot molecules.

	\paragraph{Slower-than-$O(1/N_{\text{sub}})$ scaling}
	Rather than studying VSC with a fixed number of hot molecules ($N_{\text{hot}} = 10$) and a variable number of molecules in a simulation cell ($N_{\text{sub}}$),  another approach is to keep fixed $N_{\text{hot}}/N_{\text{sub}} = 10/216 = 4.63\%$. This approach captures the physical reality that, as extensive properties,  both $N_{\text{hot}}$ and $N_{\text{sub}}$  should scale proportional to one another as a function of system size.  In Fig. \ref{fig:pbc_dependence_scaling}c, we  plot vibrational energy relaxation rates for the hot molecules with different $N_{\text{sub}}$.  Here, the outside-cavity rate (black squares) is independent of the system size. In other words, the rate is an extensive property versus the system size,  suggesting that increasing $N_{\text{hot}}$ and $N_{\text{sub}}$ at the same time is a more appropriate approach for studying the system size dependence than keeping $N_{\text{hot}} = 10$ fixed (as above).   If  we compare the inside- and outside-cavity rates,
	the average cavity effect (green stars) on vibrational energy relaxation rates remains meaningful (i.e., $> 10\%$ cavity effect compared with the rates outside the cavity) even when $N_{\text{sub}}$ reaches up to $10^4$. For example, when $N_{\text{sub}} = 8\times 10^3$, the cavity effect on the relaxation rates is $0.004$ ps$^{-1}$, which is $14\%$ of the bare relaxation rate (0.028 ps$^{-1}$) outside the cavity.

	The most interesting feature of Fig. \ref{fig:pbc_dependence_scaling}c is that the cavity effect scales with $1/N_{\text{sub}}^{0.7}$ (instead of $1/N_{\text{sub}}$); see lines with different colors which represent linear fits of the corresponding rates versus $1/N_{\text{sub}}^{0.7}$. Here, we note that  the $0.7$ in the exponent should not be regarded as a universal quantity and might vary by changing simulation parameters (e.g. the ratio $N_{\text{hot}}/N_{\text{sub}}$). 
	The underlying mechanism behind this nontrivial slower-than-$O(1/N_{\text{sub}})$ scaling comes from the opposing effects of the reduced light-matter coupling $\widetilde{\varepsilon}$ and the increased number of hot molecules ($N_{\text{hot}}$) that arises when $N_{\text{sub}}$ increases. On the one hand, when $\widetilde{\varepsilon}$ decreases proportionally to $1/\sqrt{N_{\text{sub}}}$, as mentioned below Fig. \ref{fig:pbc_dependence_scaling}a, the cavity effect on vibrational energy relaxation rates tends to exhibit an $O(1/N_{\text{sub}})$ scaling. On the other hand, according to Fig. \ref{fig:pbc_dependence_scaling}d where we plot the corresponding polaritonic spectrum for different molecular system sizes, the transiently excited LP intensifies for larger molecular systems (which is similar to Fig. \ref{fig:pbc_dependence_Nhot_only}b). Hence,  this intensified LP tends to accelerate the vibrational relaxation when $N_{\text{hot}}$ increases. Overall, these two competing effects lead to a slower-than-$O(1/N_{\text{sub}})$ scaling.

	\subsection{Polaritonic energy redistribution}
	
		\begin{figure*}[!]
		\centering
		\includegraphics[width=0.8\linewidth]{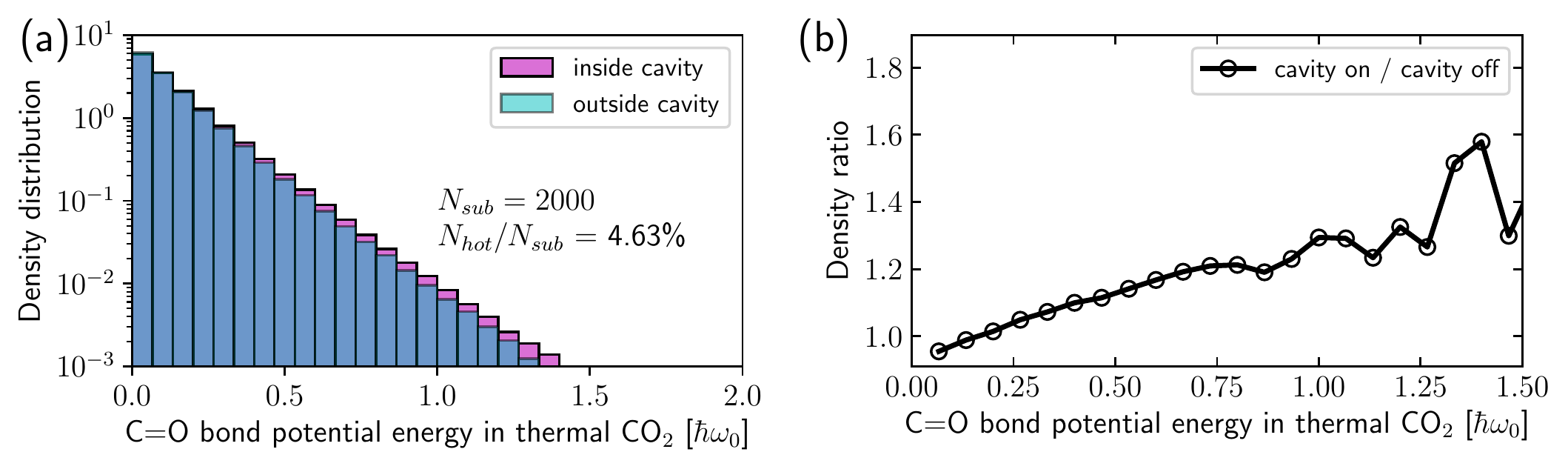}
		\caption{(a) Logarithmic-scaled density distribution of the \ch{C=O} bond potential energy as found in the molecules prepared in thermal equilibrium ("thermal molecules") during the hot-molecule relaxation process. We fix $N_{\text{sub}} = 2000$ and $N_{\text{hot}} / N_{\text{sub}} = 4.63\%$. Purple bins denote the inside cavity results; cyan bins denote the outside cavity results, which is hardly observed in the figure because the inside- and outside-cavity distributions largely overlap with each other (the overlap of purple and cyan is blue). (b) The corresponding density ratio between the inside- versus outside-cavity distribution of \ch{C=O} bond potential energy.  Note that the polaritons prefer to transfer energy to a small subset of thermal molecules located at the tail of the distribution.}
		\label{fig:nonlinear}
	\end{figure*}
	
	Another interesting and potentially significant observation is the way transiently excited polaritons redistribute their energy among molecules following the vibrational energy relaxation process. For the same conditions as in Figs. \ref{fig:pbc_dependence_scaling}c,d, Fig. \ref{fig:nonlinear}a plots the logarithmic-scaled density distribution of the \ch{C=O} bond potential energy (in unit of $\hbar\omega_0$, where $\omega_0 = 2327$ cm$^{-1}$) for the molecules prepared at thermal equilibrium (or "thermal molecules"). For this calculations, we set $N_{\text{sub}}=2000$ and run 40 NVE nonequilibrium trajectories; for each trajectory we calculate the energy distribution by taking snapshots every 1 ps during  the 40-ps simulation, so that overall we count $40\times 40 \times 2000$ \ch{CO2} configurations  during the whole relaxation and transfer process. During this time, both the outside (cyan bins) and inside (purple bins) cavity results  demonstrate an exponential distribution, which implies that the \ch{C=O} bond potential energy of thermal molecules  roughly obey a Maxwell-Boltzmann distribution; recall that the $y$-axis is on a logarithmic scale. Very interestingly, however,  the tail of the distributions of \ch{C=O} vibrational energy differ strongly  inside versus outside the cavity. This fact is more clearly shown in Fig. \ref{fig:nonlinear}b which plots the ratio of the probability density of thermal-molecule \ch{C=O} bond potential energy inside versus outside the cavity (the bins in Fig. \ref{fig:nonlinear}a).  Because both simulations start from exactly the same initial conditions and with $\widetilde{\varepsilon}$ switched on or off (see Appendix \ref{sec:simulation_details} for details), this difference in the tail distribution is a pure polaritonic effect, i.e., the transiently excited polaritons are more likely to create vibrationally higher excited molecules rather than equally distributing energy to all thermal molecules.  
	
	A possible explanation for the large difference in the tail could come from the perspective of spectral overlap between polaritons and individual molecules (which are mostly composed of vibrational dark modes). Due to anharmonicity, the molecules at the tail of energy distribution have smaller vibrational frequencies, leading to a larger spectral overlap with the LP. Therefore, the transiently excited LP would interact more strongly with the molecules at the tail and transfer more energy to these molecules than molecules with small vibrational energy.

	Finally, we remark that, since this polaritonic effect mostly takes place in the long tail of the thermal molecule \ch{C=O} vibrational energy distribution, it is possible that polariton-accelerated vibrational energy transfer may still be meaningful for a small subset of thermal molecules  even when $N_{\text{sub}}$ is very large. In other words, event though the average VSC effects per molecule will vanish once $N_{\text{sub}}$  exceeds $\sim 10^4$ (see Fig. \ref{fig:pbc_dependence_scaling}),  some molecules in the tail of the distribution  may feel the effect of the polaritons when $N_{\text{sub}}$ is beyond $\sim 10^4$ (e.g., in  Fabry--P\'erot microcavities).  Future work will investigate this possibility.

	\section{Conclusion}\label{sec:conclusion}
	We have studied the effect of VSC on vibrational energy relaxation and transfer for a small fraction of hot molecules immersed in a thermal bath of \ch{CO2} at room temperature. Several important observations have been made: (i) During this nonequilibrium process with no external pumping, polaritons, especially the LP, can be transiently excited and facilitate intermolecular vibrational energy transfer, which leads to an acceleration of vibrational energy relaxation of the hot molecules. (ii) This acceleration resonantly depends on the cavity mode detuning and can be enhanced by increasing Rabi splitting (or molecular concentration). (iii) The vibrational energy relaxation acceleration is superradiant-like and collectively scales with the number of hot molecules. (iv) For large system sizes (or large effective cavity volumes), when the fraction between the number of hot and thermal molecules remains the same, the VSC effect on the relaxation rates scales slower than $1/N_{\text{sub}}$ due to a competition between the reduced light-matter coupling ($\widetilde{\varepsilon}$) and an enhanced superradiant-like behavior of the hot molecules. (v)  Although our simulations suggest that the effect of VSC on the average relaxation rate becomes negligible when $N_{\text{sub}}$ exceeds $\sim 10^4$, polaritons are always  transiently and meaningfully excited, and the energy infused into the polaritons transfers more strongly  to the tail of  the thermal-molecule energy distribution. Altogether, this work suggests that collective VSC effects in a cavity can significantly affect vibrational relaxation energy relaxation and transfer.

	Finally, let us make a few remarks regarding the connection of this work to VSC catalytic effects on ground-state chemical reactions observed in Fabry--P\'erot microcavities. The rates of vibrational energy relaxation and transfer can significantly modify ground-state chemical reaction rates outside the cavity. For example, Kramers' theory \cite{Kramers1940,Nitzan2006} suggests that  ground-state reaction rates can depend proportionally or inversely on the energy relaxation rate. Therefore, the observation of VSC effects on  vibrational energy relaxation and transfer might imply the modification of ground-state chemical reaction rates. That being said,  VSC catalysis  is highly nontrivial as  experiments suggest (at least) the following four criteria \cite{Thomas2016,Thomas2019,Lather2019}. (i) The chemical reaction rates are modified in thermal conditions and without external polaritonic pumping. The cavity modification of chemical reaction rates (ii) resonantly depends on the cavity mode detuning and (iii) collectively depends on molecular concentration (or Rabi splitting). (iv) The cavity modification can be observed in Fabry--P\'erot microcavities (where the effective cavity volume is $\sim \lambda^3$ and $\lambda$ takes units of micrometers), meaning that the number of molecules forming VSC can reach $10^9 \sim 10^{12}$. For VSC effects on vibrational energy relaxation and transfer, we have also observed the satisfaction of criterion (i)-(iii). However, when criterion (iv) (the number limit) is considered, although we have observed a negligible average VSC effect per molecule once $N_{\text{sub}}$ exceeds $\sim 10^4$, our simulation suggests a larger polaritonic effect for molecules at the tail of the energy distribution. Because chemical reactions also occur at this same tail, such a similarity indicates that VSC effects on vibrational energy relaxation and transfer could possibly play a significant role in VSC catalysis --- a hypothetical premise that deserves further study.

	\section{Acknowledgments}
	This material is based upon work supported by the U.S. National Science Foundation under Grant No. CHE1953701 (A.N.); 
	and US Department of Energy, Office of Science,
	Basic Energy Sciences, Chemical Sciences, Geosciences, and Biosciences Division (J.E.S.).

	
	\begin{appendices}
		
		\setcounter{equation}{0}
		\setcounter{figure}{0}
		\setcounter{table}{0}
		\renewcommand{\theequation}{S\arabic{equation}}
		\renewcommand{\thefigure}{S\arabic{figure}}
		\renewcommand{\bibnumfmt}[1]{[S#1]}
		\setcounter{section}{0}
		\renewcommand{\thesection}{S-\Roman{section}}

		\section{Method}\label{sec:method}
		
		CavMD propagates the following equations of motion for the coupled photon-nuclei system:
		\begin{subequations}\label{eq:EOM_MD_PBC}
			\begin{align}
				M_{nj}\ddot{\mathbf{R}}_{nj} &= \mathbf{F}_{nj}^{(0)}  + \mathbf{F}_{nj}^{\text{cav}}
				\\
				m_{k,\lambda}\ddot{\dbtilde{q}}_{k,\lambda} &= - m_{k,\lambda}\omega_{k,\lambda}^2 \dbtilde{q}_{k,\lambda}
				-\widetilde{\varepsilon}_{k,\lambda} \sum_{n=1}^{N_{\text{sub}}}d_{ng,\lambda}
			\end{align}
		\end{subequations}
		Here, subscript $nj$ denotes the $j$-th nucleus in molecular $n$, $M_{nj}$, $\mathbf{R}_{nj}$, $\mathbf{F}_{nj}^{(0)}$, and $\mathbf{F}_{nj}^{\text{cav}}$ denote the mass, position, nuclear force, and cavity force for nucleus $nj$; see Refs. \cite{Li2020Water,Li2020Nonlinear} for the exact forms of the forces.  Subscripts $k,\lambda$ denote the cavity photon mode with wave vector $k$ and polarization direction $\lambda =x, y$ for a $z$-oriented cavity (see Fig. \ref{fig:toc} for the simulation setup), $m_{k,\lambda}$, $\dbtilde{q}_{k,\lambda}$, $\omega_{k,\lambda}$ denote the auxiliary mass, position, and frequency of cavity photon mode $k,\lambda$. The cavity photon mode $k,\lambda$ interacts with the dipole moments of molecules ($d_{ng,\lambda}$) with an effective coupling strength $\widetilde{\varepsilon}_{k,\lambda}$. 
		
		\subsection{System size dependence}
		
		One important feature of CavMD  is the use of periodic boundary conditions. In detail,  cavity photons interact with $N_{\text{cell}}$ identical simulation cells, each of which contains $N_{\text{sub}}$ molecules, so the total molecular number becomes $N = N_{\text{sub}}N_{\text{cell}}$. The replica of $N_{\text{cell}}$ simulation cells has been reflected in the definition of 
		\begin{equation}
			\widetilde{\varepsilon}_{k,\lambda} = \sqrt{N_{\text{cell}}} \varepsilon_{k,\lambda}
		\end{equation}
		in Eq. \eqref{eq:EOM_MD_PBC}, where $\varepsilon_{k,\lambda}$ is the true coupling strength between a single molecule and the cavity mode $k,\lambda$. 
		
		When studying how VSC effects can depend on the molecular system size (or the molecular number), we can take $N_{\text{cell}} = 1$ and study the molecular response for different choices of $N_{\text{sub}}$ while keeping the molecular density and Rabi splitting the same. Here, the Rabi splitting is unchanged if we modify $\widetilde{\varepsilon}_{k,\lambda}$ according to $\widetilde{\varepsilon}_{k,\lambda} \propto \sqrt{1/N_{\text{sub}}}$. The corresponding CavMD results will reflect the VSC response for a liquid system in cavities with  the same polaritonic frequencies but with different effective volumes \footnote{Note that periodic boundary conditions are always applied to exclude the edge effect of the simulation cell.}.
		
		When this system size dependence is studied, according to a second order perturbation calculation,  the VSC effect on individual molecules should scale as $O(\widetilde{\varepsilon}_{k,\lambda}^2) = O(1/N_{\text{sub}})$. This $O(1/N_{\text{sub}})$ scaling will quickly remove any VSC effects on individual molecules once $N_{\text{sub}}$ is large enough; as has been noted earlier \cite{Pilar2020}, this approach cannot explain any collective cavity effect.
		
		\section{Simulation details}\label{sec:simulation_details}
		
		We simulate a model yet realistic molecular system under VSC: the \ch{C=O} asymmetric stretch mode (peaked at 2327 cm$^{-1}$) of  a large ensemble of liquid-phase \ch{CO2} molecules forms collective VSC with a near resonant cavity mode. The detailed procedure to perform CavMD simulations for such a liquid \ch{CO2} system is given in Ref. \cite{Li2020Nonlinear} and all input files to generate results in this manuscript are available at Github \cite{TELi2020Github}. Therefore, below we only briefly outline the simulation details. 
		
		As shown in Fig. \ref{fig:toc}, the simulation setup consists of $N_{\text{sub}} = 216$ \ch{CO2} molecules in a cubic simulation cell (with cell length 24.292 \si{\angstrom}, which corresponds to a molecular density 1.101 $\text{g/cm}^{3}$) confined within a pair of metallic mirrors along the $z$-direction. An anharmonic force field \cite{Li2020Nonlinear} is used to propagate the dynamics of \ch{CO2}. During the simulation, only a single cavity photon mode (with two polarization directions: $x$ and $y$) is considered and the effective coupling strength is set as $\widetilde{\varepsilon} = 2\times 10^{-4}$ a.u..
		
		We are interested in  the cavity modification of both the vibrational energy relaxation of the hot \ch{CO2} molecules and the subsequent  intermolecular vibrational energy transfer to the thermal \ch{CO2} molecules. Before simulating this nonequilibrium dynamics, outside a cavity ($\widetilde{\varepsilon}=0$), we first run 150 ps NVT  (constant molecular number, volume, and temperature) simulations at 300 K to equilibrate the system and then run 40 consecutive NVE (constant molecular number, volume, and energy) trajectories with duration 20 ps. Starting from the initial configurations (with both position and velocity information) of the above 40 equilibrium NVE  trajectories outside a cavity ($\widetilde{\varepsilon}=0$), we prepare a nonequilibrium initial condition from equilibrium configurations  by resampling the initial velocities of $N_{\text{hot}}=10$ arbitrary \ch{CO2} molecules (in total $N_{\text{sub}} =216$ \ch{CO2} molecules are simulated). The initial conditions of the hot molecules are reset so that their kinetic energy in each degree of freedom  obeys a uniform distribution in an interval $6000 \pm 250$ K. 
		Such a random resampling of velocities has been chosen to mimic the preparation of uncorrelated hot molecules in a thermal bath at room temperature,  where the effective temperature of these hot molecules would be $\sim 3000$ K (since the initial positions of the hot molecules are not modified  and still obey a thermal distribution).

		Starting from each of the 40 nonequilibrium initial conditions, we then run nonequilibrium NVE simulations for 40 ps and calculate physical properties outside the cavity ($\widetilde{\varepsilon}=0$) by averaging over the 40 nonequilibrium trajectories. Inside the cavity, we start from exactly the same  nonequilibrium initial configurations as the outside cavity case but reset  $\widetilde{\varepsilon} = 2\times 10^{-4}$ a.u. to switch on the light-matter coupling and run NVE trajectories. Note that the use of the NVE ensemble implies that we have neglected any cavity loss, which simplifies the interpretation of results.

		\section{On calculating polaritonic spectrum}\label{Appendix:spectrum}
		
		Because polaritons are composed of  a molecular bright mode and cavity photons, a polaritonic spectrum can be obtained from either the molecular or the photonic side.
		From the molecular side, a polaritonic spectrum can be obtained by calculating the molecular infrared (IR) absorption spectrum, which can be evaluated by Fourier transforming the dipole auto-correlation function \cite{McQuarrie1976,Gaigeot2003,Habershon2008,Nitzan2006}:
		\begin{equation}\label{eq:IR_equation_cavity}
			\begin{aligned}
				n(\omega)\alpha(\omega) &= \frac{\pi \beta \omega^2}{2\epsilon_0 V c} \frac{1}{2\pi}   \int_{-\infty}^{+\infty} dt \ e^{-i\omega t}  \\
				& \times \avg{\sum_{i=x, y}\left(\vmu_S(0)\cdot \ve_i\right) \left(\vmu_S(t)\cdot \ve_i\right)} 
			\end{aligned}
		\end{equation}
		Here, $\alpha(\omega)$ denotes the absorption coefficient, $n(\omega)$ denotes  the refractive index, $\beta= \kB T$, $V$ is the volume of the system (i.e., the simulation cell),
		$\ve_i$ denotes the unit vector along direction $i=x, y$, and $\vmu_S(t)$ denotes the total dipole moment of the molecules at time $t$. Fig. \ref{fig:cavity_effect}a and Fig. \ref{fig:detuning_dependence}a are calculated by evaluating Eq. \eqref{eq:IR_equation_cavity} from equilibrium NVE trajectories.
		
		Similar as Eq. \eqref{eq:IR_equation_cavity}, in order to obtain the polaritonic spectrum, we can also define a photonic coordinate auto-correlation function:
		\begin{equation}\label{eq:ph_equation_cavity}
			\begin{aligned}
				n(\omega)\alpha_{k}(\omega) \propto \omega^2   \int_{-\infty}^{+\infty} dt \ e^{-i\omega t}   \avg{\sum_{\lambda=x, y}\dbtilde{q}_{k,\lambda}(0)\dbtilde{q}_{k,\lambda}(t)} 
			\end{aligned}
		\end{equation}
		where $\alpha_{k}(\omega)$ denotes the absorption coefficient for cavity photon mode $k$. 
		Fig. \ref{fig:pbc_dependence_Nhot_only}b and Figs. \ref{fig:pbc_dependence_scaling}b,d are calculated by evaluating Eq. \eqref{eq:ph_equation_cavity} from nonequilibrium NVE trajectories  during the whole simulation period (40 ps). Note that when nonequilibrium trajectories are used to calculate Eq. \eqref{eq:ph_equation_cavity}, the resulting spectrum is a transient spectrum which reflects the average dynamic behavior of photons during the nonequilibrium trajectories.

	\end{appendices}

	
	%

\end{document}